# Design for the Right to the Smart City in More-than-Human Worlds

**Sara Heitlinger**
Newcastle University
Newcastle-upon-Tyne, UK
sara.heitlinger@newcastle.ac.uk

**Rob Comber**
Royal Institute of Technology
Stockholm, Sweden
rcomber@kth.se

## ABSTRACT
Environmental concerns have driven an interest in sustainable smart cities, through the monitoring and optimisation of networked infrastructure processes. At the same time, there are concerns about who these interventions and services are for, and who benefits. HCI researchers and designers interested in civic life have started to call for the democratisation of urban space through resistance and political action to challenge state and corporate claims. This paper aims to add to the growing body of critical and civic-led smart city literature in HCI by leveraging concepts from the environmental humanities about "more-than-human" worlds, as a way to shift understandings within HCI of smart cities away from the exceptional and human-centered, towards a more inclusive understanding that incorporates and designs for other "others" and other species. We illustrate through a case study that involved co-designing Internet of Things with urban agricultural communities, possibilities for creating more environmentally and socially just smart cities.

## INTRODUCTION
As cities become fertile grounds for embedded IoT technologies and services, HCI has become increasingly interested in the projects, visions and narratives of their integration in what has become known as the smart city. In particular, environmental concerns have driven an interest in sustainable smart cities, through the optimisation of urban processes and resources, services and infrastructures, making them more efficient and therefore more sustainable [28]. The building of eco-cities is now at the "forefront of national and global agendas" [51]. Typical examples involve the use of networked sensing and tracking technologies for low-carbon infrastructure (e.g. smart energy metering, to reduce waste and emissions [44], water recycling and automated collection systems [51] and increasing efficiency in food supply chains [29].

HCI researchers and designers interested in civic life have started to question "Whose right to the smart city?" recalling French philosopher Henri Lefebvre's call for the democratisation of urban space through resistance and political action, amid growing concerns about who these interventions and networked infrastructures are for and who benefits [5]. Critics have voiced concerns over: who owns, controls, and has access to proprietary smart city infrastructure [2,26,28]; privacy, surveillance and censorship [2,19]; inequalities in terms of representation, participation, and access [2,64]; and the encroachment of algorithmic culture into government, civics and public life [19,28]. "More and more commentators these days critique the established hegemony of the engineering and technology-centric epistemology embedded in any one proprietary smart city vision." [26].

Another critique of these visions is the way in which they are tackling the problem of urban sustainability, by increasing productivity whilst achieving efficiency [28]. There are 3 main critiques of this approach: 1) Like other modernist, top-down, efficiency-based, techno-solutions to the problem of sustainability that have already been critiqued within HCI [11,61], eco-smart cities are subject to particular types of breakdown, because they are unable to deal with the complexities of real, messy cities [51]. 2) Sustainability gets performed in a specific way that leaves little room for political participation or citizen agency [23,28,51]. 3) There are critiques over approaches to sustainability that merge economic growth with green objectives [46] inherent in smart cities narratives [28], as economic growth is seen as the cause of environmental degradation [50], and we should, therefore, be aiming instead for "prosperous descent" [1].

One thing that both the socio-political and the environmental critiques share is a human-centered perspective of cities, in which urban space, separated from nature, is designed, planned for, and used by human inhabitants only. A human exceptionalist perspective is increasingly seen as problematic within the social sciences and humanities, due to the environmental destruction, climate change, and loss of species that are implicated. In these disciplines, there is a turn towards a more complex understanding of the ways in which human-environmental relations are enmeshed. Recent work in design has embraced these alternative understandings to explore the ways in which humans and other species are interrelated in cities, and with technology [19,25,60]. We build on this emerging work in design and HCI by using an expanded ontology of cities with which to explore the right to the smart city and begin to address the socio-political and environmental critiques levelled against their dominant narratives. For, as Houston has argued, "any presumed exclusive human 'right to the city' and the biosphere is increasingly untenable" [40].

We situate our work within smart cities, civic IoT, and sustainable HCI to make the following contributions: First, we introduce Lefebvre's notion of the "right to the city", and explore the ways in which designers are participating in the radical struggle to reclaim urban space, both digital and physical – or "hybrid" when taken together – from the forces of capitalism and the state. Secondly, we draw on the "more-than-human" literature from STS, the environmental humanities and cultural and urban geography to propose a productive lens to investigate the question of the right to the smart city in multi-species worlds. We present a case study that involved co-designing networked environmental sensors, data visualisation, and a seed library for urban agricultural communities as an example of a design research project in this space. And finally, we present a critical analysis of this study through different rights to the smart city, reflecting on the role of design and HCI in creating socially and environmentally just hybrid space in which humans and non-human others co-produce, cohabit and co-manage urban life-worlds.

## RIGHT TO THE CITY

The "right to the city" formulation that French philosopher Henri Lefebvre coined in 1968 [48], is a declaration of a collective intention to struggle against homogenizing planetary urbanization. It is a commitment to become active and move towards the democratization of urban space, to reappropriate the production of space from the dominant hegemonic regimes, which in contemporary cities, is neo-liberalism. Within neoliberalism, space is valued predominantly for its exchange value, and private property and profit is prioritised over all other rights and claims [33,47,57]. Lefebvre gives the following example: "There are two ways in which urban space tends to be sliced up, degraded and eventually destroyed by this contradictory process: the proliferation of fast roads and of places to park and garage cars, and their corollary, a reduction of tree-lined streets, green spaces, and parks and gardens. The contradiction lies, then, in the clash between a consumption of space which produces surplus value and one which produces only enjoyment – and is therefore 'unproductive'. It is a clash, in other words, between capitalist 'utilizers' and community 'users' [47]

For Lefebvre, rights are not codified protections guaranteed by the state, achievements that come at the end of a struggle like the US Civil Rights Act of 1964 [57]. Rather they come at the beginning, and are political declarations of an intention to struggle. Historically rooted in revolution and in Marxism, "Rights are people voicing their commitment to become active and to move together in a particular direction, towards a particular horizon" [57].

Within the dominant existing neoliberal capitalist system, the production of space is alienated, or made strange, from the users, because it is not produced by them, but by others for them. We are also alienated from others who share the space. The right to the city can be articulated in terms of the right to *spatial autogestion*, which refers to the radical project of people refusing to passively accept the existing system of spatial production, one of property rights on which the capitalist economy exists. To counter this alienation people must reappropriate the production of space, take control of it and govern it for themselves [57].

Top-down decision-making processes turn urban locations into abstract spaces, where people are also alienated from each other. Spatial autogestion returns those spaces back into specific places. "Spatial autogestion reverses the separation and segregation of inhabitants; it draws them together into common spaces where they would encounter each other and engage in meaningful discussions about the city and its future."[57]

The acts of spatial autogestion are happening continuously in our cities: people and social movements everywhere, in all manner of ways, are engaged in active struggle to reshape the city and overcome isolations, resisting the efforts of developers, and the state to create homogenising urban space for capitalist and state benefit [33]. Lefebvre (according to [57]) says that if we want to participate in the right to the city, then we must identify the sites of struggle, learn to see them, narrate them, and help them proliferate. Rather than focus on the structures of power, it is more productive to spend our energy cultivating the world we want to live in by participating, narrating and sharing it.

### Right to the Smart City

With the proliferation of networked sensing and digital infrastructures into urban life, questions over rights to the city are now being raised in relation to the smart city [2,5,19,26]. There are concerns that the algorithms that drive these technologies, and the data produced will be steered towards increasing profits of huge companies, rather than towards increasing civic participation. "Unlike the physical urban space that it overlays, this new and rapidly emerging "virtual" space has practically no capacity constraints. However, it is subject to inequalities in terms of access, representation, participation, and ownership. Indeed, today it is mostly large corporations like Google, Facebook and Twitter that control the digital social interactions at a global scale…the complete lack of ownership and control of these platforms on the users' behalf poses significant threats related to privacy, surveillance, censorship, and manipulation, which should not be underestimated" [2]. There are concerns over urban citizenship "reduced to a series of actions focused on monitoring and managing data, when that data managed by corporate and state actors" [28], as well as the creep of "algorithmic culture into government, civics, and our public lives" [19].

As [2] has argued, "These concerns raise the issue of the citizens' right to the digital city, and if both the physical and virtual are considered together, the "right to the hybrid city." Furthermore, "there is a gap today between those that fight for our rights to the city with those that fight for our rights to ICTs, despite the fact that in the times of the smart city, these two objectives are more and more interwoven" (ibid). This interweaving is exemplified in Apple's new "town squares", where communities are encouraged to form around Apple's products in hybrid space [14].

The struggle for the production of hybrid space is of interest to HCI, in its concerns over control, participation, representation, ownership, access, surveillance, and privacy as well as attempts by civic-minded technology efforts to appropriate the production of that space from corporate and state interests. For example, [5] explored how citizens from disadvantaged backgrounds can "participate in the collection, sharing and use of data to tackle issues of their own concern", for example through participatory sensing. [2,5,27] advocate for a digital commons, in which city infrastructure, services, and data – such as urban sensing, or local DIY WiFi networks, are "appropriated at the grassroots level and for the common good" [5], and help empower citizens to claim their rights to privacy, freedom of expression, diversity, and self-determination [2]. Internet of things technologies is used for civic media and as a way to address "matters of concern and care." In [19] While [64] are helping citizens gain the knowledge and skills to make use of, or to make sense of data, thereby overcoming barriers of access and participation.

Documented in this literature, then, are examples of digital spatial autogestion, in which citizens are appropriating the means of production of the smart city for themselves, thereby countering the alienation caused by spatial practices that regulate it for the benefit of corporations or the state. Rather than passively accepting or consuming the existing system of spatial production in the smart city, people are taking up the challenge of understanding and mastering that production for themselves.

This right to the hybrid smart city has been otherwise described as smart citizenship, in which smart citizens resist technocratic determinism of the smart city through bottom-up, community-driven, low-cost, and local innovative efforts such as "open access and user-centered systems in which the smart use of information can increase transparency, accountability, participation, and collaboration" [13].

Community groups are also addressing sustainability within the smart city, but rather than focusing on efficiency, such efforts take into account a bottom-up environmental citizenship and agency [51] which go beyond merely acting as a node in a cybernetic city [28] for example through pollution monitoring [5], citizen sensing of nature [19], cycling [3]. These examples add to the ways in which designers are participating in the struggle for the democratisation of digital space by supporting citizen-led infrastructuring, civic IoT, data access and literacy, and commoning, by facilitating between communities and smart technologies in relation to matters of concern, and by documenting and narrating these sites of struggle in the design and HCI literature, therefore helping them to proliferate.

## MORE-THAN-HUMAN CITIES

We now come to a discussion of the "more-than-human" literature from cultural and urban geography, science and technology studies, and environmental humanities, to show how it could be productively used within smart cities discourses and visions. We draw on very recent work in both design and HCI that has started to touch on these themes, in response to environmental concerns.

### Ontological Exceptionalism of Humanism in Cities

Cities are often understood as separate from nature, and are primarily planned and built for human inhabitants [40,60]. This is because we typically understand the 'urban' in a human exceptionalist way. Human exceptionalism is similar to urban exceptionalism and stems from traditions within Western knowledge to think in hyper-separated categories, or dualisms: e.g., human/non-human; nature/culture; mind/body; city/nature etc.). Within human exceptionalism the human is a separate, autonomous individual, superior to the non-human, living in a sovereign body whose actions do not have ecological consequences [54]. Cities may have been built out of natural materials but they are now "elevated to places of progressive human and technological mastery" [40]. Within this way of thinking, "human relations are not only distinct from nature, but are effectively independent of the web of life" [50]. This kind of thinking allows us to see other species as resources to be exploited, or as a nuisance to be eliminated, for the higher human needs.

Within the social sciences and humanities, in fields such as STS, geography, and now the environmental humanities, a common thread that runs through this literature is that human exceptionalism and other kinds of dualistic thinking such as Nature/Society no longer serve us in the age of the Anthropocene. This term refers to a proposed new geological era in which human activity is transforming earth systems [40], accelerating climate change and causing mass extinctions [50]. Within a Human/Nature binary, the "living, multi-species connections of humanity-in-nature" (ibid) are converted into dead abstractions that "connect to each other as cascades of consequences rather than constitutive relations" (ibid). Within cities, human exceptionalism has resulted in "asymmetric 'negotiations' between human planners and nonhuman others", which have contributed to environmental destruction [40].

Preferring the term Capitolocene to acknowledge the role of capital in the current age of environmental destruction, Moore argues that the Nature/Society dualism obscure "our vistas of power, production and profit…. It prevents us from seeing the accumulation of capital as a powerful web of interspecies dependencies; it prevents us from seeing how those interdependencies are not only shaped by capital, but also shape it; and it prevents us from seeing how the terms of that producer/product relation change over time." For these reasons it is difficult for us to see how, and adapt to, the ways in which "capitalogenic climate change is undermining crucial relations of capitalism's Cheap Food regime in the twenty-first century – Cheap Nature increasingly confronts forms of nature that cannot be controlled by capitalist technology or rationality [50]. "Capitalism's governing conceit is that it may do with Nature as it pleases, that Nature is external and may be fragmented, quantified and rationalized to serve economic growth, social development or some other higher good. This is capitalism as a project." (ibid), but this results in environmental destruction. For example, soil around the world is being degraded due to intensive farming that capitalises on short term profits over long-term sustainability [63]. Within the Capitalocene, the same system that causes social injustice results in environmental destruction, because people and natural resources are exploited for capital.

### Multi-Species Entanglements

Within this literature, the Anthropocene is a cause for rethinking binaries such as Human/Nature, a step towards asking questions about the inseparability of humans and nature [25,32,60], and as a call for expanding our understanding of the entanglements between human and nonhuman worlds. This type of thinking is not new in HCI. For example, drawing on STS, HCI has often understood the world to be made of up "hybrids, assemblages, and collectives that are composed of human and nonhumans that act and organize together, sharing the delegation of power and agency as understood by actor-network theory" [40] and object oriented ontology [8,43] and nonanthropocentric design [22,59]. Within design and HCI too, there are those who advocate for design to conceive of humans and animals in a relational perspective, as a way to overcome problematic narratives of human privilege and exceptionalism. [25,60,62].

A perspective in which we understand the imbricated nature of humans and nature, humans and non-humans, and cities and nature, can have important implications for how we think about environmentally sustainable cities. For we are, in fact, "only one species among many inhabiting diverse urban worlds" [12], And despite these cultural narratives, "city dwellers are deeply entangled with natural elements, including plant life, animals, dirt, water" [60]. We must, therefore, rethink human-environment relationships in terms of the complex biophysical worlds that we inhabit [17]. Houston argues that those involved in the planning and building of urban space – which includes HCI researchers and designers working in the smart city space – must consider human/non-human and human/environmental relations if we want to create environmentally and socially just cities "in a time of global environmental uncertainty and change" [40].

In the remainder of this paper we present a design research case study through which we explore the concerns and lenses discussed above.

## CASE STUDY: CONNECTED SEEDS AND SENSORS

Connected Seeds and Sensors is a design research project that sought to better understand how IoT can support urban sustainability, particularly in the context of food production and consumption in the smart city. It explored co-designing digital and networked technologies and the development of an interactive artefact, the Connected Seeds Library, with urban agricultural communities in east London. The research aimed to expand the design space of sustainable smart cities beyond managerial, utilitarian and efficiency-based narratives by incorporating playful interactions, personal stories, and co-creation with urban farmers. The research proposal was developed together with Anon City Farm, with which we have a long-standing collaboration, described in [34,35,37,38]. The seed library artefact illustrates our community engagement and design research process, now lives permanently at the farm, and serves as a resource to local communities.

### Anon City Farm

Like many community gardens in the UK, Anon City Farm started in the 1970s by a group of local people who occupied vacant land to grow fresh food. Food is grown all-year round in rotation, with seeds being planted to replace the food that will soon be finished. It has a diverse base of volunteers and visitors in terms of age, ability, socio-economic background and ethnicity. Through its community gardens, volunteer opportunities, its various educational programmes and fresh produce sales, the farm encourages local communities to grow and consume healthy fresh food.

As discussed in [36–38], and like the other gardens in east London that were involved in the project, Anon Farm places a high value on inclusivity, education and health and well-being of people and the Earth. Its environmental work not only includes food-growing and healthy eating activities, but also involves capacity building by strengthening knowledge and skills within nearby communities, and an integrative approach to the management of food production and waste cycles.

The farm and the other urban agricultural sites we engaged are located in the east London boroughs of Tower Hamlets, Hackney and Walthamstowe, which are some of the most deprived economically in the UK. Tower Hamlets in particular, where the farm is located, has been characterised by high population density, large-scale immigration, ethnic diversity, and poverty and huge divides between rich and poor. At the same time, Tower Hamlets also contains Canary Wharf, one of London's two main financial centres and home to some of the world's largest banks. It has proportionally more people earning above £90,000 and more earning less £15,000 than the London average, and the gap between the two extremes is growing [68]. The farm supports Somalian, Zimbabwean, Bengali and Turkish community gardening groups, as well as school kids, people suffering from post- traumatic stress and mental health service users. There are high levels of racial segregation in the Borough with around 50% of secondary schools being entirely non-white. It has the highest number of school pupils in England whose first language is not English (74%)[67], and it has the highest rate of child poverty across the UK [69]. It suffers from a range of food-related illnesses, for example, adults in the borough are more likely to have diabetes compared to the rest of London and England and one of the highest childhood obesity levels in the country. As Bagwell [4] has argued, food security is a concern in urban centres, particularly in deprived neighbourhoods with large ethnic minority populations because access to affordable healthy food may be compounded by cultural or religious dietary requirements. In Tower Hamlets, 76% of households are within a 10-minute walk of a supermarket, but 97% are within a 10-minute walk of a fast food outlet [15].

## From Talking Plants to Connected Seeds

Connected Seeds and Sensors was an 18-month project that took place between October 2015 and March 2017, funded through the UK Engineering and Physical Sciences Research Council's Internet of Things – Research in the Wild stream. The proposal was developed collaboratively with Anon City Farm, and researchers at Queen Mary University of London, and grew out of the Talking Plants [38], a ludic engagement with living edible plants, that contributed to the educational and community engagement work of the farm. Through this work we identified opportunities for internet of things technologies to support the practices of food-growing and seed-saving in east London, which is particularly rich due to the ethnic diversity of the region. The Connected Seeds Library was designed to collect and share this knowledge, connect people to their heritage through food, and to make available locally-grown seeds that have adapted to local climates, and which may be of unusual or heritage varieties that are not available in commercial catalogues. It was designed as an engaging, interactive and playful artefact, that does not require any particular technical skill, or ownership of smartphone or computer.

The research team were interested in exploring networked environmental sensors, typically used to optimize food production through resource management, pest control, and waste reduction in precision agriculture, within the context of small-scale urban farming. Through a participatory design methodology, we wanted to include more diverse voices into the debate about what sustainability means in the smart city, and how such understandings can influence the design space. By focusing on seed-saving and urban agriculture we sought to better understand the interrelations between the environment, health and wellbeing, economics, politics, and the social elements that impact on sustainability.

We were interested in seeds as a vehicle to explore the role of IoT in sustainable cities and in particular regarding the complexities of seed-sovereignty (referring to the control of seed production and supply), biodiversity, community-based agriculture and the city. These concerns were pressing when we developed the proposal due to proposals to change EU law that would make it mandatory to register seed (at substantial cost), which campaigners claimed would be disastrous to biodiversity, farmers' rights, and play into the hands of big business such as Monsanto [66] .

## What We Did

### *Requirements gathering workshops*

We started with a series of workshops with growers in east London, aimed to better understand the needs, practices and values of small-scale and community urban growers and seed-savers. From the data generated in these workshops we began to form what we called "data categories" which related to the information our growers would find interesting or useful to know about seeds. This included practical advice such as when to plant seeds, but also cultural and personal associations and recipes.

### *Engaging seed guardians*

From February-October 2016 we recruited and engaged 20 seed guardians who committed to grow 1-2 crops for seed, and to donate some of those seeds to the library at the end of the season. They also agreed to take photos of their gardens, plants, harvests and meals cooked. We recorded audio interviews with them at the start of the season, and again at the end, which were structured around the different qualitative data categories that we had elicited in the workshop phase. Of the twenty that initially signed up to the project, fourteen remained engaged till the end. Seed guardians were culturally diverse, with origins from Bangladesh, the Caribbean, Turkey, Zimbabwe, France, Belgium, Britain, Australia, China, and Ireland). They had differing levels of gardening experience from complete beginner, to professionally qualified. Four of the gardeners were growing their plants at Anon Farm on different communally-managed plots and raised beds; one was growing on a separate community garden; six were growing on public land on two housing estates; and three were growing in their private gardens. Of those who began, but then left the project without donating any seeds, two were on a "meanwhile garden", a community garden on land temporarily allocated by developers until they were ready to build, and another had a rooftop garden. Crops grown for seed included: kodu, lablab beans, Zimbabwean pumpkin, Zimbabwean maize, calaloo, orach, summer purslane, chickpeas, pak choi, Thai basil, runner beans, achocha, black mustard, chili, tomato, winter squash, red Russian kale, coriander and long beans.

We organised community events throughout the growing season, including seed-swaps, seed-saving workshops, garden visits and design sessions. The aim of these activities was to support a community of practice, where seed guardians could share skills and knowledge, maintain motivation, and participate in design decisions about the Connected Seeds Library. Additional engagement was through a documentary film, a book, an exhibition, and a day of festive events.

*Sensors*

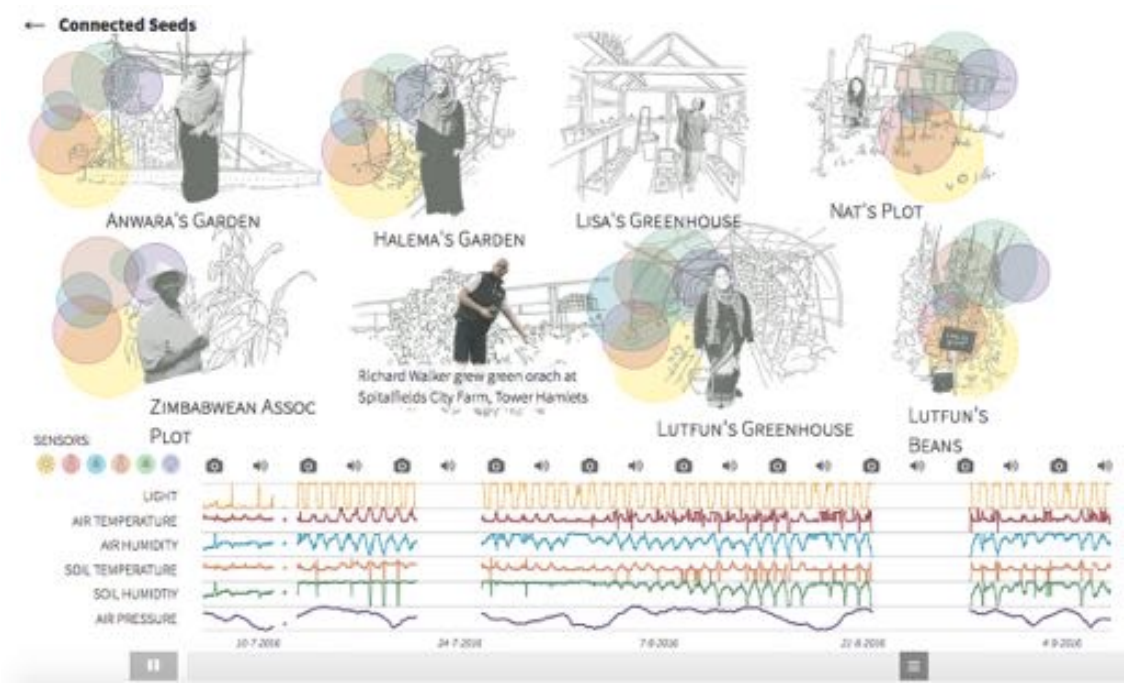

**Figure 1. Screen grab of data visualisation**

At the same time, we installed custom-built networked environmental sensors in eight of the gardens. These collected information about air temperature, air humidity, air pressure, soil moisture, soil temperature, and ambient light. The system was built by an engineer in the research team, who used an Arduino board, with high precision sensors embedded in a purpose-built packaging in order to protect the sensors, the battery pack, and the board from potential rain or moisture in the environment. A reading was taken from the sensors every hour, and, in order to save on battery life, sent once a day to a web server over a 3G network, as WiFi was not available in the gardens. We collected data for 2 months.

At the end of the growing season we hired a data visualisation company to present the data collected in an interactive webpage. In addition to the quantifiable data produced from the sensors – represented by a graphical animation across a timeline – photographic images pop up and audio clips from the Seed Guardians begin to play as the timeline progresses [70].

*The Connected Seeds Library*

The Connected Seeds Library is the design artefact that was produced with the help of a professional artist called Franc Purg. The library contains seeds donated by the Seed Guardians, as well as the digital records (audio and photographs) connected to those seeds. Visitors to the library can select a jar of seeds and place it on a designated pad in order to start a slideshow of images connected to that seed. They can turn a wheel to play audio tracks of the Seed Guardian talking about their experiences of growing. There are ten different categories on the wheel for the audio: Who I am; Why I grow my own food; Why I save my own seeds; Connections to my heritage; How I feel when I'm working in the garden; Where these seeds came from; How I grow them; Tips and tricks for growing these seeds; Recipes; How to save seeds. These headings we established from the workshops. The audio comes from the interviews, which were edited into 1-2-minute-long tracks.

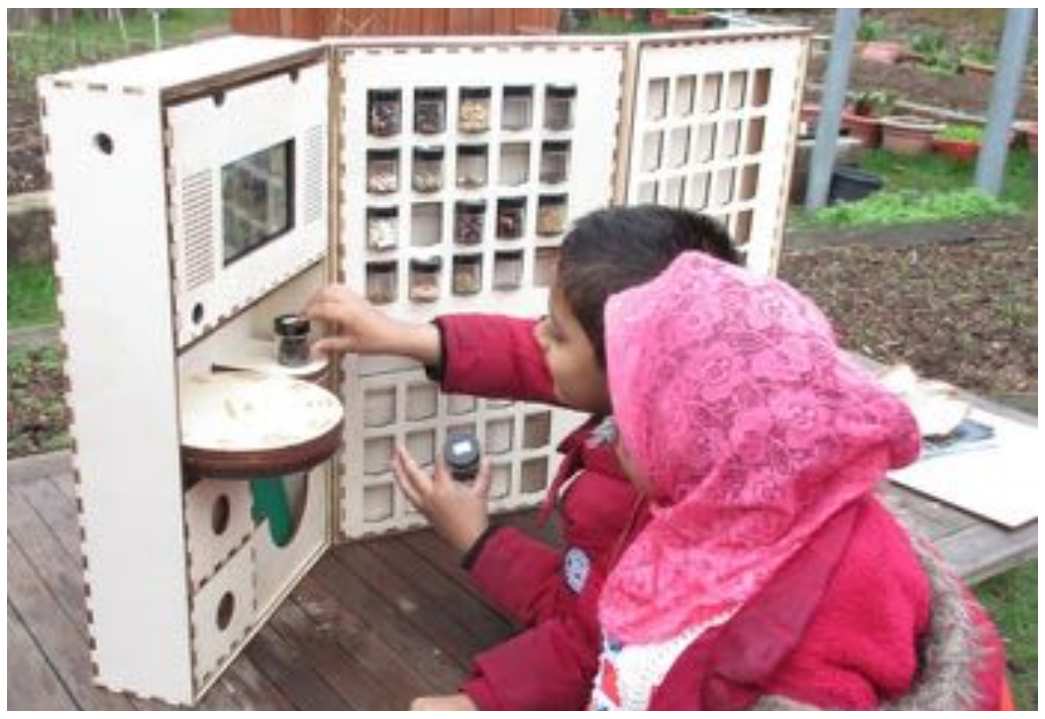

**Figure 2. Engaging with the Connected Seeds Library**

People are invited to join the library for free, take some seeds to grow at home, and bring some back at the end of the season to maintain the living stock.

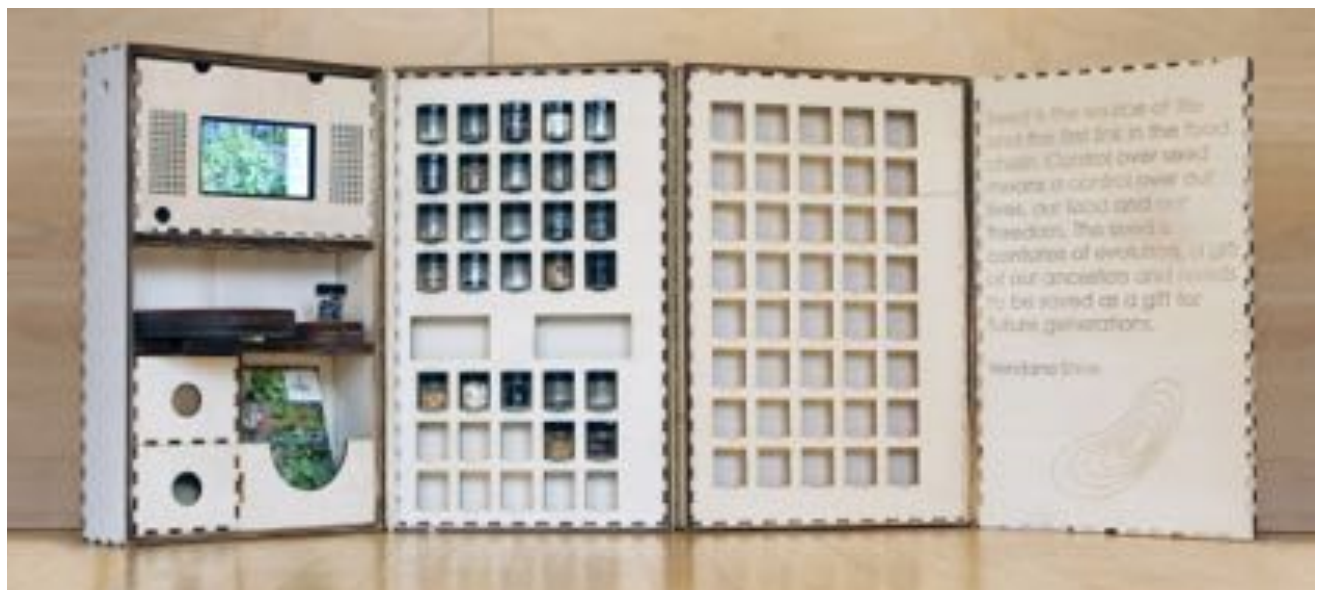

**Figure 3. The Connected Seeds Library**

The interactive elements include a Raspberry Pi and screen, an RFID reader, an Arduino microcontroller, a battery, and a speaker. The seed jars are tagged with RFID tags. The library also contains seed packets with QR codes and web links to the project website (connectedseeds.org), and specifically to the page about those specific seeds grown by that guardian.

The Connected Seed Library continues to be used as a community resource at its permanent home at Anon Farm.

## DISCUSSION

In this section, we discuss what it means to talk of design for the right to the hybrid smart city through an expanded more-than-human perspective, as articulated through the following: the right to difference; the right to care time; and the right to the commons.

**The Right to Difference**

For Lefebvre, capitalism turns the city into homogenous, abstract space that denies differences. In this way, it segregates urban inhabitants and alienates them from each other in an effort to produce passive consumers instead of active citizens. Likewise, neoliberalism turns the smart city into abstract space through global and non-specific internet infrastructures [2], which alienates citizens into passive consumers, or reduces them to passive nodes in a cybernetic system [28]. Design often perpetuates the visions of sustainable smart cities as full of "Resource Men", white, middle-class technofetishists, "cast in the image of the male-dominated industries of engineering and economics that permeate energy management" [61] and who dream them into being. Within these visions, if there are any people at all, they are alienated from the production of space, from nature, from each other, and from the production of food. Communities of colour are often marginalised, so the right to the city is enmeshed in the struggles against marginalization [57]. The multi-ethnic neighourboods that the Connected Seeds Library serves are impacted by increasing land prices as well as food deserts and food-related health conditions. Within City/Nature divides, this marginalisation extends to other species, who may be seen as pests to be exterminated, or as non-existent [30,50].

The practices of spatial autogestion reverse the process of alienation by creating "sites of encounter" [57] with difference, with "other" people and other species. By drawing people together into common spaces, urban agricultural sites provide places to break down social segregation, overcome racism and other kinds of exceptionalism that are on the rise in the UK and elsewhere, and thereby help contribute to social cohesion. They dismantle narratives of human exceptionalism and urban exceptionalism, based on binaries such as City/Nature and Human/Non-Human by bringing different diverse peoples and species together literally and figuratively in hot steaming compost piles [32], where they can interact, play, share experiences, and practice multispecies care. In this way, the cultivation of urban land is one example of the ways in which everyday practices of urban inhabitants enact the struggle for the right to the city.

The Connected Seeds Library uses smart city technologies of IoT, networked sensing, and data visualisation to augment, amplify, narrate and celebrate the ways in which diverse people and species are brought together in urban agricultural sites. Interacting with the seed library tells the following story of diversity: "We really love to come here and meet other people from our home. And other people from other places. We introduce them to our crops and we see their crops here. Some are similar to our crops but not the same." (Basilia). It tells of mutual caring and sharing: "I keep the seeds and I keep plants living their whole life for the animal biodiversity so there's insects coming in and the birds eat the seeds. So, there's enough for everyone" (Kate). And it tells of reversing alienation from others in urban space: "When I came to this country it was very difficult for me. All the neighbours spoke English, just me, Bengali. I was scared and thought how can I go outside and meet people and talk to them? After twenty years I started gardening. Now I'm not scared of anything. Gardening changed my life" (Anwara). The seed library participates in autogestion by narrating and supporting the spatial practice of gardening that bring diverse people, species and cultures together in the web of life.

There may be competing needs in the right to the city, which may need to be negotiated. For example, different community groups may compete for space to grow produce from their different cultural backgrounds and places of origin. At the same time, different species may compete for life and death [32]. Weeds, slugs and snails are done away with, as are other "awkward creatures" [30]; while ladybirds are encouraged because they control aphids, bees and hoverflies are encouraged because they support pollination. The seed library acknowledges this negotiation and killing [10], and narrates this complexity, without seeking a quick fix: "The biggest thing that's going to impact on the slugs and snails is things like the blackbirds and the thrushes, so make sure you've got lots of shrubbery, and keep a pond for frogs. It's not like an immediate cure, but it's a sustainable long-term cure, and it makes life a lot easier to correct that imbalance in the biodiversity, rather than intervening too much" (Richard). By narrating these stories, the seed library requires of us to consider complex interspecies entanglements in the production of space. It teaches us to stay with the trouble, as Haraway urges us, not reverting to visions of technofixes embodied in top-down sustainable smart cities narratives, nor succumbing to fatalism [32]. As the seed guardian explains: "Gardening is… a process. It's about life. It's about growth and death and decay, which is all together" (Ahmet). "Eschewing futurism, staying with the trouble is both more serious and more lively. Staying with the trouble requires making oddkin; that is, we require each other in unexpected collaborations and combinations, in hot compost piles. We become-with each other or not at all. That kind of material semiotics is always situated, someplace and not noplace, entangled and worldly"[32].

**The Right to Care Time (in Care Space)**

Top-down visions of sustainable smart cities that aim for efficiency and optimisation perform a distinct version of sustainability that is removed from environmental citizenship, and that merge "green" objectives with economic growth, an idea that is inherently contradictory in the Capitolocene, which sees capitalism as the cause of the environmental crises [50].

The Anthropocene asks of us to consider different time scales in designing the right to the smart city, for example, “care time” suggested by Maria de la Bellacasa [55]. The Connected Seeds Library engages with care time as rhythmical, and asks us to consider the growth of plants and seed in relation to the seasons of the year, e.g. when to plant, when to replant outside, when to harvest, when to save seeds. It demonstrates how city time and city labour can be recuperated from capitalism through the telling of stories from seed guardians, for example about the mental health benefits of gardening: “It gives headspace. Especially in an urban environment where there is very little space for people. And time. It is all about space and time to think” (Debbie). Through the slow process of gardening, the gardeners enact the right to the city, as a right to “change ourselves by changing the city more after our heart’s desire” (Harvey). Design must be careful not to do away with practices of care time, for example by automating gardening practices or compromising opportunities for face to face interactions as argued previously in [7,37,52].

Dominant narratives of the Capitalocene such as Cheap Nature and Cheap Food [50], have intensified the rhythms of agricultural production and perceive soil as a “receptacle for crops” [55], resulting in 1.3 billion people “trapped on degrading agricultural land” [24]. Rather than putting a networked sensor in the soil to extract data for increased productivity and efficiency, the project engages with soil “as a living community” (ibid). The soil sensor data taken together with the human stories tell of mutually caring human-soil relations that progress over care time. For example, one guardian, used the sensor data to validate his climate-adaptive gardening practices that involved mulching the soil to support a moist and fertile environment, without the accessing mains water. In another example, a seed guardian who is a counsellor working with immigrant communities who have endured torture, tells of the psychological benefit of caring for the soil: “Being active, doing things outside, working in the soil, with soil, these things can change moods easily.” (Ahmet). The idea that engagement with the microbes in soil is good for mental health is scientifically supported. Slowing down allows for practices of mutual interspecies care to flourish, and counters the narratives of the Capitolocene.

The seed library also deals with the timescales of memory, in which knowledge, tradition, and cultural practices of growing and preparing food, and saving seeds, are passed down through the generations. Seed Guardians speak of how they’ve turned the clock back to slower, less “efficient” practices from their parents’ generations because they recognise that supporting biodiversity, seed-saving, and practices that nourish the soil should be valued in the long term. “Modern agriculture came in and slightly devastated all traditional farming methods. The concept of saving seed went out the window. I've gone back to more organic production and traditional farming and gardening methods, which is the kind of the difference between me and my parents' generation. In the 70's and on they were like "Get out the old, don't need it anymore." And now I'm putting it all back.” (Kate). We suggest that the visions of the smart city based on the “predominant timescales of technoscientific futurity and their reductive notion of innovation” [55] are no longer viable with the pace required by ecological multi-species care.

### The Right to the Commons

Capitalism values urban resources such as space for is exchange value, rather than its use value, privileging “capitalist utilizers” over “community users” [47]. Urban agriculture prioritises the collective needs of inhabitants (human and otherwise) over individual property rights, and urban resources such as seeds, soil, worms, flowers, compost and tools, are kept in the commons. The commons refers to “commonly held property, use, stewardship and management of the available and produced resources” [2] by a community. As hybrid digital-physical space becomes increasingly important for the lives of city dwellers, having impact on both social inclusion and the natural environment, there are increasingly strong demands for the digital resources such as data, networks, location based services, sensors and other devices to be appropriated for the urban digital commons [27]. Citizens should make collective decisions how to management that infrastructure and the rights of all inhabitants to participate in the governance of the smart city must be considered not just a private elite (ibid).

HCI is working for the right to the city commons. For example, Balestrini et al. (2016) have developed a city commons framework for designers to support “citizens, especially those from disadvantaged communities, [to] participate in the collection, sharing and use of data to tackle issues of their own concern, including noise pollution, housing conditions, or social isolation” [5]. In order to be able to participate in the urban digital commons, citizens will need the technical skills and data literacy in order to negotiate data ownership and governance (ibid), which is currently recognised as a barrier to participation [65]. While [19] have been exploring how digital sensing technologies may contribute to commoning, through shared activities and knowledge.

What does the digital urban commons mean within “more-than-human” worlds? Seeds and growers are connected in their struggle for seed sovereignty, which refers to the ownership and control of seed, which is under threat by large agribusinesses who want to control seed [41]. The seed library participates in the struggle for the urban commons by keeping the seeds in the urban commons, available for free as a community resource. Furthermore, these seeds have been locally grown, and many are of unusual or heritage varieties that are not available in seed catalogues, and in this way, contribute to seed diversity in light of the dwindling number of seed varieties available in commercial seed catalogues. Seed guardians talk of the role that the seed library plays in seed sovereignty: "Over the years lots of seed varieties and heirloom seeds have been lost in favour of commercially grown crops. So, for me seed sovereignty is about taking the control back and being able to collect our own seeds and being able to carry on doing practices that farmers have been doing for a long time all over the world" (Nat) "You know, seeds, food, water, the air that we breath, they're all basic human rights, and I think business needs to back off… [The seed library] is just that nudge. Why do we get sucked into that whole commercialism, you get the glossy seed catalogue and January, February time you're poring over it and planning, and we should just be sharing what we've grown."? (Debbie).

Rather than seeing seed as a commodity whose value is produced through exchange, the seed library gives each seed the ability to participate in the urban commons. It has unique and important role, and a story to tell in the web of life, one that is entangled with the stories of life, death, culture, migration, land, climate, power and politics. The seed library collects these seeds and the stories of the seeds and demonstrates the ways in which we are all interrelated and interdependent. By providing a home for those seeds, making them available as a free resource to other people, along with stories of their growers and the environmental conditions under which they were grown, the seed library participates in the production of hybrid space, and wrests back control and management of the vested interests of capital, and also from state and corporate interests who want to control the land for private property. The seed library tells the story of the seed, and tells of the importance of the seed, and of the importance of seed sovereignty. At the same time, practical and cultural knowledge of growing and preparing those seeds becomes part of the urban commons. By sharing the seeds and the stories about those seeds the seed library is helping to maintain the urban commons through seed sovereignty.

The rights of property and profit are individual rights, while the right to the city and the urban commons is a collective right. Harvey [33] asks us to consider what sort of city we want to live in, and in asking this question, we are actually asking "what kind of people we want to be, what kinds of social relations we seek, what relations to nature we cherish, what style of daily life we desire, what kinds of technologies we deem appropriate" (ibid). By bringing people together in urban agricultural sites, and growing a network of people who can share seeds and knowledge of growing in the city, the seed library amplifies the spatial autogestion beyond the individual human, plant, or garden. Through the skill-sharing, seed-swapping activities that were part of the project and continue to this day, the seed library contributes to community building, bringing people together to encounter difference, gain inspiration and discuss how they could productively manage urban space. We envisage that this collective ownership of the commons could be strengthen further through a hybrid digital/physical network of seed libraries that serve different communities and locations, and in this way to address scale [23], while at the same time being sensitive to local socio-ecological contexts.

At the same time, ownership of the urban commons bears a collective responsibility: "You can't just have a seed library and then it be done. You've got to keep it running: so, who's going to grow this seed this year, so that we can have fresh seed next year, and it's from this location. All of that is really important" (Kate). Drawing attention to the collective responsibility asks us to consider the sustainability of Cheap Food and Cheap Nature in the age of the Capitolocene [50], and the collective struggle [33] for the right to the city: We should be "expanding the network and making [the seed library] a place that everyone feels they can go to. And also, food security, you know it's getting hotter, [food is] more expensive already. We've all got to work harder in our communities, about making food growing and sharing of food more viable." (Kate). Seed guardians draw attention to the collective responsibility of maintaining the urban commons, in the Anthropocene.

In addition, the project participates in the digital commons by using open-source hardware and software, which is available for others to download and use [45]. The seed library counters the abstract digital spaces of smart cities [2]. The technology used is not part of generic multi-national infrastructure, but specific and locally owned, generated and maintained and for the benefit of the multi-species communities. The data, both from the sensor data and from the stories and images are available for all to see on the project website.

However, there remain hard questions to answer about how to make sensor data meaningful to communities, even if they own it. Although there were some interesting responses from the guardians, the sensor data was largely opaque to both researchers and guardians. Typically it is designers and other "experts" who decide how the knowledge is generated and "who has the capacity to contribute towards addressing climate change" [59]. It is certainly not trivial to make ecological data accessible to the general public, but examples from art [42,58,59] suggest productive ways forward.

## THE ROLE OF DESIGN

Rather than focus on the hegemonic structures of power, or retreat into one or other of the twin illusions of techno-progress or fatalism [32], we urge designers to take up the call to struggle for the right to the hybrid smart city in more-than-human worlds. For, as Houston et. al. have argued, "any presumed exclusive human 'right to the city' and the biosphere is increasingly untenable" [40]. Design can play an important role in this struggle by identifying sites where spatial autogestion is already taking place, strengthening and amplifying their work through design practice, and narrating and sharing the process through design research, thereby helping them grow and proliferate. This asks of us to expand our thinking of the smart city in terms of human exceptionalism and human-centered design, for "socio-technical systems are the site of politics, values, and ethics where cities are being made" [25].

As designers, we make choices about where we put our time and energy. We could decide to invest our efforts in designing smart cities infrastructure for cars, which intensify the production of space in particular ways and for particular bodies and interests (e.g. car manufacturers and oil companies). A human exceptionalist lens would fail to see the ways in which other people (e.g. non-drivers) and other species are marginalized from this production of space. Alternatively, we could choose to design for multi-species flourishing in care time. For example, one seed guardian suggested a citywide infrastructure for networked pollen monitoring, that would allow gardeners to collectively coordinate the planting of pollinator-friendly plants that account for seasonal timescales. "If there's a particular month when there's really low pollen, we'd need to think about things that flower at that time and try and fill in the gaps, for bees and other pollinators" (Nat). This idea could build on previous work in HCI on pollen sensing [21] but expand the focus beyond human benefit.

Design is well placed to do this work, and is able to draw on existing and well documented methodologies to assist in this task. Speculative participatory design [6,25], ludic participatory design [34], design fictions, speculative civics [20], object oriented ontologies, and multi-species ethnographies [53] to name just a few may be useful in decentering the human in design to consider the roles and perspectives of non-human others in smart cities. Design can draw on theories from STS and the environmental humanities to provide theoretical lenses through which to understand the socio-political-ecological-technical-ethical complexities and help focus endeavours. Design artefacts can raise provocative questions, dilemmas and possibilities for multi-species spatial practices to perform autogestion in hybrid digital-physical space, and to demonstrate productive collaborations in which humans and nonhuman actors cohabit, co-produce, and co-manage the urban commons, in ways that are respectful of difference and in timescales that are more nourishing of our relations and our Earth. For example, the right to difference in the smart city through a more-than-human lens asks us to consider design that may have little benefit to humans, none at all, or may even be at the expense of humans, for example [16]. In designing for care time, design can take inspiration from other slow practices such as Slow Technology [31], Slow Food, and Soil Time [56] that offer alternatives to rationalistic and efficiency-led ways of being in, making, and understanding the world.

## CONCLUSION

Within HCI we are starting to see alternatives to the visions of the top-down, managerial, efficiency-led smart city, in which citizens participate, access, govern, and own the digital commons. At the same time, we can see that HCI, drawing on fields such as STS and the environmental humanities is beginning to consider the Anthropocene as a lens through which to respond to global environmental concerns, including climate change, loss of biodiversity, increasing food prices and pollution of air, water and land. In this paper, we have built on this emerging work to address the socio-political and environmental critiques levelled against the dominant narratives and visions of smart cities, by employing an expanded ontology of cities in the critical analysis of the Connected Seeds design research case study.

Most often, design is understood as disconnected from the politics of consumption [23]. But as HCI starts to take the Anthropocene seriously, as Light et al. [49] has argued passionately it can no longer be business as usual. Design and HCI cannot separate itself from politics even if it wishes to, because they don't exist in a political vacuum. Design is informed by the cultural narratives we tell ourselves – of natural resources being unlimited, of human exceptionalism, of technological progress, of the unstoppable nature of free market capitalism, and of the incompatibility of nature and agriculture with urban space. The current global economic, humanitarian, and environmental crises demand a change in these cultural narratives [11,18].

Focusing on sites of struggle for the right to the city, such as urban agricultural communities, highlights the ways in which we can begin to shift the conceptualisation of sustainability within HCI from a discourse of sustainable consumption [39] towards a collective, participatory and holistic understanding that takes into account social and environmental justice within multispecies contemporary urban life. It allows us to expand the design space of smart cities beyond human and urban exceptionalism, towards seeing the environmental crisis as a communal problem that requires communal action, where individuals can work collectively in the city to ameliorate the destructive impact of our current practices on the environment [9]. By working with such sites, we can observe changes in cultural and political narratives in action. We recognise that grassroots urban agricultural communities, and other communities where the struggle for the right to the city is ongoing, are not separate from the capitalist system within which they function, and therefore they still participate in it. But because of their values, and the inclusive practices in care time and space, which help strengthen the links between collective action, participation and environmental citizenship, they present a site where such shifts can begin to occur. Design has an important role to play in supporting and strengthening these shifts in the hybrid smart city.

## REFERENCES

1. Samuel Alexander. 2015. *Prosperous descent: Crisis*

*as opportunity in an age of limits*. Simplicity Institute.

2. Panayotis Antoniadis, Ileana Apostol, Mark Gaved, Michael Smyth, and Andreas Unteidig. 2015. DIY networking as a facilitator for interdisciplinary research on the hybrid city. *Hybrid Cities 2015: Data to the People*: 65–72.
3. Mariam Asad and Christopher A Le Dantec. 2017. Tap the: A Design-Based Inquiry into Issue Advocacy and Digital Civics. In *Proceedings of the 2017 CHI Conference on Human Factors in Computing Systems*, 6304–6316.
4. Susan Bagwell. 2011. The role of independent fast-food outlets in obesogenic environments: a case study of East London in the UK. *Environment and Planning A* 43, 9: 2217–2236. https://doi.org/10.1068/a44110
5. Mara Balestrini, Yvonne Rogers, Carolyn Hassan, Javi Creus, Martha King, Paul Marshall, Knowle West, and Media Centre. 2017. A City in Common : A Framework to Orchestrate Large - scale Citizen Engagement around Urban Issues. 2282–2294.
6. Karl Baumann, Benjamin Stokes, François Bar, and Ben Caldwell. 2017. Infrastructures of the Imagination : Community Design for Speculative Urban Technologies. 15–18. https://doi.org/10.1145/3083671.3083700
7. Eric P S Baumer and M Six Silberman. 2011. When the Implication is Not to Design (Technology). In *Proceedings of the SIGCHI Conference on Human Factors in Computing Systems* (CHI '11), 2271–2274. https://doi.org/10.1145/1978942.1979275
8. Ian Bogost. 2012. *Alien phenomenology, or, what it's like to be a thing*. U of Minnesota Press.
9. A Boucher, D Cameron, and N Jarvis. 2012. power to the people: dynamic energy management through communal cooperation. *Proceedings of the Designing* ... 70: 612–620. https://doi.org/10.1145/2317956.2318048
10. Jeremy Brice. 2014. Killing in More-than-human Spaces : Pasteurisation , Fungi , and the Metabolic Lives of Wine. 4: 171–194.
11. Hronn Brynjarsdottir, Maria Håkansson, James Pierce, Eric Baumer, Carl DiSalvo, and Phoebe Sengers. 2012. Sustainably unpersuaded: How Persuasion Narrows Our Vision of Sustainability. *Proceedings of the 2012 ACM annual conference on Human Factors in Computing Systems - CHI '12*: 947. https://doi.org/10.1145/2207676.2208539
12. Jason Byrne. 2010. The human relationship with nature. *The Routledge Handbook of Urban Ecology. London: Routledge*: 63–73.
13. Igor Calzada, Cristobal Cobo, Igor Calzada, and Cristobal Cobo. 2016. Unplugging : Deconstructing the Smart City Unplugging : Deconstructing the Smart City. 732, January. https://doi.org/10.1080/10630732.2014.971535
14. Kelsey Campbell-Dollaghan. 2017. Stores Are Not Town Squares. Retrieved September 17, 2017 from https://www.fastcodesign.com/90139799/stores-are-not-town-squares
15. Martin Caraher, S Lloyd, and T Madelin. 2009. *Cheap as chicken: fast food outlets in Tower Hamlets (report no 2)*.
16. Tristan Cork. 2017. Street lights on Bristol to Bath cycle path switched off so glow worms can find love. Retrieved September 18, 2017 from http://www.bristolpost.co.uk/news/local-news/street-lights-bristol-bath-cycle-182389
17. Aidan Davison. 2015. Beyond the mirrored horizon: modern ontology and amodern possibilities in the Anthropocene. *Geographical Research* 53, 3: 298–305.
18. Carl Disalvo, Kirsten Boehner, and Nicholas A Knouf. 2009. Nourishing the ground for sustainable HCI: Considerations from ecologically engaged art. *Proceedings of the 27th International Conference on Human Factors in Computing Systems (CHI '09)*: 385–394. https://doi.org/http://doi.acm.org/10.1145/1518701.1518763
19. Carl Disalvo and Tom Jenkins. 2017. Fruit Are Heavy : A Prototype Public IoT System to Support Urban Foraging. 541–553.
20. Carl DiSalvo, Tom Jenkins, and Thomas Lodato. 2016. Designing Speculative Civics. In *Proceedings of the 2016 CHI Conference on Human Factors in Computing Systems*, 4979–4990.
21. Carl Disalvo, Thomas Lodato, Jonathan Lukens, and Tanyoung Kim. 2014. Making Public Things : How HCI Design Can Express Matters of Concern. 2397–2406.
22. Carl DiSalvo and Jonathan Lukens. 2011. Nonanthropocentrism and the nonhuman in design: possibilities for designing new forms of engagement with and through technology. In *From social butterfly to engaged citizen: urban informatics, social media, ubiquitous computing, and mobile technology to support citizen engagement*. 440–460.
23. Paul Dourish. 2010. HCI and environmental sustainability: the politics of design and the design of politics. *Proceedings of the 8th ACM Conference on Designing Interactive Systems . ACM.*: 1–10. https://doi.org/10.1145/1858171.1858173
24. Nigel Dudley and Sasha Alexander. 2017. *Global Land Outlook*. Retrieved from https://global-land-outlook.squarespace.com/the-outlook/#the-bokk
25. Laura Forlano. 2016. Decentering the Human in the

Design of Collaborative Cities. 32, 3. https://doi.org/10.1162/DESI

26. Marcus Foth. 2015. *Citizen ' s Right to the Digital City*. https://doi.org/10.1007/978-981-287-919-6

27. David Franquesa and Leandro Navarro. 2017. Sustainability and participation in the digital commons. *interactions* 24, 3: 66–69.

28. Jennifer Gabrys. 2014. Programming environments : environmentality and citizen sensing in the smart city. 32: 30–48. https://doi.org/10.1068/d16812

29. Filippo Gandino, Bartolomeo Montrucchio, Maurizio Rebaudengo, and Erwing R Sanchez. 2009. On improving automation by integrating RFID in the traceability management of the agri-food sector. *IEEE Transactions on Industrial Electronics* 56, 7: 2357–2365.

30. Franklin Ginn, Uli Beisel, and Maan Barua. 2014. Flourishing with awkward creatures: Togetherness, vulnerability, killing. *Environmental Humanities* 4, 1: 113–123.

31. Lars Hallnäs and Johan Redström. 2001. Slow technology - designing for reflection. *Personal and Ubiquitous Computing* 5, 3: 201–212. https://doi.org/10.1007/PL00000019

32. Donna J Haraway. 2016. *Staying with the trouble: Making kin in the Chthulucene*. Duke University Press.

33. David Harvey. 2008. THE RIGHT TO THE CITY. 1–16.

34. Sara Heitlinger. 2016. Talking Plants and a Bug Hotel: Participatory Design of ludic encounters with an urban farming community. Queen Mary University of London.

35. Sara Heitlinger and Nick Bryan-Kinns. 2013. Understanding performative behaviour within content-rich Digital Live Art. *Digital Creativity* 24, 2: 111–118. https://doi.org/10.1080/14626268.2013.808962

36. Sara Heitlinger and Nick Bryan-Kinns. 2013. Understanding performative behaviour within content-rich Digital Live Art. *Digital Creativity* 24, 2: 111–118. https://doi.org/10.1080/14626268.2013.808962

37. Sara Heitlinger, Nick Bryan-Kinns, and Janis Jefferies. 2013. Sustainable HCI for grassroots urban food-growing communities. In *Proceedings of the 25th Australian Computer-Human Interaction Conference: Augmentation, Application, Innovation, Collaboration*, 255–264.

38. Sara Heitlinger, Nick Bryan-Kinns, and Janis Jefferies. 2014. The talking plants: an interactive system for grassroots urban food-growing communities. In *CHI'14 Extended Abstracts on Human Factors in Computing Systems*, 459–462.

39. K. Hobson. 2002. Competing Discourses of Sustainable Consumption: Does the “Rationalisation of Lifestyles” Make Sense? *Environmental Politics* 11, 2: 95–120. https://doi.org/10.1080/714000601

40. Donna Houston, Diana Maccallum, Wendy Steele, and Jason Byrne. 2017. Make kin , not cities ! Multispecies entanglements and “ becoming-world ” in planning theory. https://doi.org/10.1177/1473095216688042

41. Dan Iles. 2017. Seed Sovereignty. *Connected Seeds*.

42. Rachel Jacobs, Steve Benford, Mark Selby, Michael Golembewski, Dominic Price, and Gabriella Giannachi. 2013. A Conversation Between Trees: What Data Feels Like in the Forest. In *Proceedings of the SIGCHI Conference on Human Factors in Computing Systems* (CHI '13), 129–138. https://doi.org/10.1145/2470654.2470673

43. Tom Jenkins, Christopher A Le Dantec, Carl DiSalvo, Thomas Lodato, and Mariam Asad. 2016. Object-oriented publics. In *Proceedings of the 2016 CHI Conference on Human Factors in Computing Systems*, 827–839.

44. Yoshifuyu Karakasa, Hirohiko Suwa, and Toshimizu Ohta. 2007. Evaluating Effects of RFID Introduction Based on CO2 Reduction. In *Proceedings of the 51st Annual Meeting of the ISSS-2007, Tokyo, Japan*.

45. Nanda Khaorapapong. 2016. Connected Seeds sensors. Retrieved September 17, 2017 from https://github.com/haddadi/cs-sensor-unit

46. Bran Knowles, Lynne Blair, Mike Hazas, and Stuart Walker. 2013. Exploring Sustainability Research in Computing: Where We Are and Where We Go Next. In *Proceedings of the 2013 ACM International Joint Conference on Pervasive and Ubiquitous Computing* (UbiComp '13), 305–314. https://doi.org/10.1145/2493432.2493474

47. Henri Lefebvre. 1991. *The production of space*. Oxford Blackwell.

48. Henri Lefebvre. 1996. The right to the city. *Writings on cities* 63181.

49. Ann Light, Irina Shklovski, and Alison Powell. 2017. Design for Existential Crisis. In *Proceedings of the 2017 CHI Conference Extended Abstracts on Human Factors in Computing Systems*, 722–734.

50. Jason W Moore. 2017. The Capitalocene, Part I: on the nature and origins of our ecological crisis. *The Journal of Peasant Studies* 44, 3: 594–630. https://doi.org/10.1080/03066150.2016.1235036

51. Paul D Mullins. 2017. The Ubiquitous-Eco-City of Songdo : An Urban Systems Perspective on South Korea ' s Green City Approach. 2, 2: 4–12. https://doi.org/10.17645/up.v2i2.933

52. William Odom. 2010. “Mate, We Don’T Need a Chip to Tell Us the Soil’s Dry”: Opportunities for Designing Interactive Systems to Support Urban Food Production. In *Proceedings of the 8th ACM Conference on Designing Interactive Systems* (DIS ’10), 232–235. https://doi.org/10.1145/1858171.1858211

53. Hannah Pitt. 2015. On showing and being shown plants - a guide to methods for more-than-human geography. 48–55. https://doi.org/10.1111/area.12145

54. Val Plumwood. 2010. Nature in the active voice. *Climate change and philosophy: transformational possibilities*: 32–47.

55. Maria Puig de la Bellacasa. 2015. Making time for soil: Technoscientific futurity and the pace of care. *Social Studies of Science* 45, 5: 691–716. https://doi.org/10.1177/0306312715599851

56. Maria Puig de la Bellacasa. 2017. *Matters of Care: Speculative Ethics in More Than Human Worlds*. University of Minnesota Press, Minneapolis.

57. Mark Purcell and Shannon K Tyman. 2017. Cultivating food as a right to the city. 9839, July. https://doi.org/10.1080/13549839.2014.903236

58. Alexandra Regan Toland. 2016. Dust Blooms. Retrieved September 19, 2017 from https://artoland.wordpress.com/2016/06/28/dust-blooms/

59. Nancy Smith and Jeffrey Bardzell. 2017. Designing for Cohabitation : Naturecultures , Hybrids , and Decentering the Human in Design. 1714–1725.

60. Nancy Smith, Shaowen; Bardzell, and Jeffrey Bardzell. 2017. Designing for Cohabitation : Naturecultures , Hybrids , and Decentering the Human in Design. *Chi*: 1714–1725. https://doi.org/http://dx.doi.org/10.1145/3025453.3025948

61. Yolande Strengers. 2014. Smart Energy in Everyday Life: Are you Designing for Resource Man? *Interactions* 21, 4: 24–31. https://doi.org/10.1145/2621931

62. Alex S Taylor. 2017. What Lines, Rats, and Sheep Can Tell Us. *Design Issues*.

63. Jonathan Watts. Third of Earth’s soil is acutely degraded due to agriculture. Retrieved from https://www.theguardian.com/environment/2017/sep/12/third-of-earths-soil-acutely-degraded-due-to-agriculture-study

64. Annika Wolff. 2017. Creating an Understanding of Data Literacy for a Data-driven Society Creating an Understanding of Data Literacy for a Data-driven Society.

65. Annika Wolff, Milton Keynes, Matthew Barker, and Marian Petre. Creating a Datascape : a game to support communities in using open data. https://doi.org/10.1145/3083671.3083686

66. Campaign for Seed-Sovereignty. Retrieved September 18, 2017 from http://www.seed-sovereignty.org/EN/

67. 2012. *EAL pupils in primary and secondary schools by LEA 2004-2013*. Retrieved from http://www.naldic.org.uk/research-and-information/eal-statistics/eal-pupils

68. 2013. Household income in Tower Hamlets. Retrieved September 18, 2017 from https://www.towerhamlets.gov.uk/Documents/Borough_statistics/Income_poverty_and_welfare/Research-Briefing-2013-04-Household-Income-final.pdf

69. 2016. Poverty in your area 2016. Retrieved from http://www.endchildpoverty.org.uk/poverty-in-your-area-2016/

70. 2017. Connected Seeds visualisation. Retrieved September 18, 2017 from http://www.connectedseeds.org/data-visualisation/